\documentclass[]{hdsr}

\graphicspath{{figs/}}

\begin{document}

\newgeometry{bottom=1.5in}

\volumeheader{0}{0}{00.000}

\begin{center}

  \title{Designing for interactive exploratory data analysis requires theories of graphical inference}
  \maketitle

  \thispagestyle{empty}
  
  \vspace*{.2in}

  \begin{tabular}{cc}
    Jessica Hullman\upstairs{\affilone,*}, Andrew Gelman\upstairs{\affilone}
   \\[0.25ex]
   {\small \upstairs{\affilone} Department of Computer Science, Northwestern University.} \\
   {\small \upstairs{\affiltwo} Department of Statistics and Department of Political Science, Columbia University, New York.} 
  \end{tabular}
  
  \emails{
    \upstairs{*}jhullman@northwestern.edu 
    }
  \vspace*{0.4in}

\begin{abstract}
Research and development in computer science and statistics have produced increasingly sophisticated software interfaces for interactive and exploratory analysis, optimized for easy pattern finding and data exposure. But design philosophies that emphasize exploration over other phases of analysis risk confusing a need for flexibility with a conclusion that exploratory visual analysis is inherently ``model free'' and cannot be formalized. 
We describe how without a grounding in theories of human statistical inference, research in exploratory visual analysis can lead to contradictory interface objectives and representations of uncertainty that can discourage users from drawing valid inferences.
We discuss how the concept of a model check in a Bayesian statistical framework unites exploratory and confirmatory analysis, and how this understanding relates to other proposed theories of graphical inference. Viewing interactive analysis as driven by model checks suggests new directions for software and empirical research around exploratory and visual analysis. For example, systems might enable specifying and explicitly comparing data to null and other reference distributions and better representations of uncertainty. Implications of Bayesian and other theories of graphical inference can be tested against outcomes of interactive analysis by people to drive theory development.    
\end{abstract}
\end{center}

\vspace*{0.15in}
\hspace{10pt}
  
\copyrightnotice

\section*{Media Summary}
Novel interactive graphical user interface tools for exploratory visual data analysis provide analysts with impressive flexibility in how to look at and interact with data. Often these systems are designed to make patterns in data as easy to see as possible. However, there are risks to prioritizing easy finding of patterns alone as a criteria of good interface design. One risk is that the techniques used to emphasize patterns, like aggregating data by default, cause analysts to overlook variation and uncertainty in their data, so that they draw conclusions that aren't well supported by the data. Another is that some analysts may fail to recognize the importance of doing more careful statistical modeling to investigate how reliable insights they arrive at through visual search seem to be. One reason that graphical user interface systems for interactive analysis may not be designed to enforce strong connections between exploratory and confirmatory statistical analysis is because there aren't well-established theories of how these two types of activities are related. We propose a perspective that unites exploratory and confirmatory analysis through the idea of graphs as model checks in a Bayesian statistical framework, and describe how in light of this view, systems for exploratory visual analysis should be designed to better support model-driven inference and representation of uncertainty.

\restoregeometry
\newgeometry{bottom=0.5in}

\section{What is the role of data visualization in hypothesis-driven analysis?}


``Nothing---not the careful logic of mathematics, not statistical models and theories, not the awesome arithmetic power of modern computers---nothing can substitute here for the flexibility of the informed human mind,'' wrote \citet{tukey1966data} half a century ago. Since then, research areas like information visualization and interactive analytics have become thriving subfields of computer science, motivated by an assumption that interactive visual interfaces for querying data enable humans to combine their domain knowledge with data summaries to produce insight. 
This has led to the development of interactive interfaces to help analysts more easily conduct ad hoc data exploration and analysis, from programmatic environments like computational notebooks, to modern business intelligence tools that create dashboards or trellis plots without the user needing to manually specify encodings, to visualization recommenders that serve up data summaries optimized for perception and exposure of patterns.

\begin{figure*}
\centering
 \includegraphics[width=\textwidth]{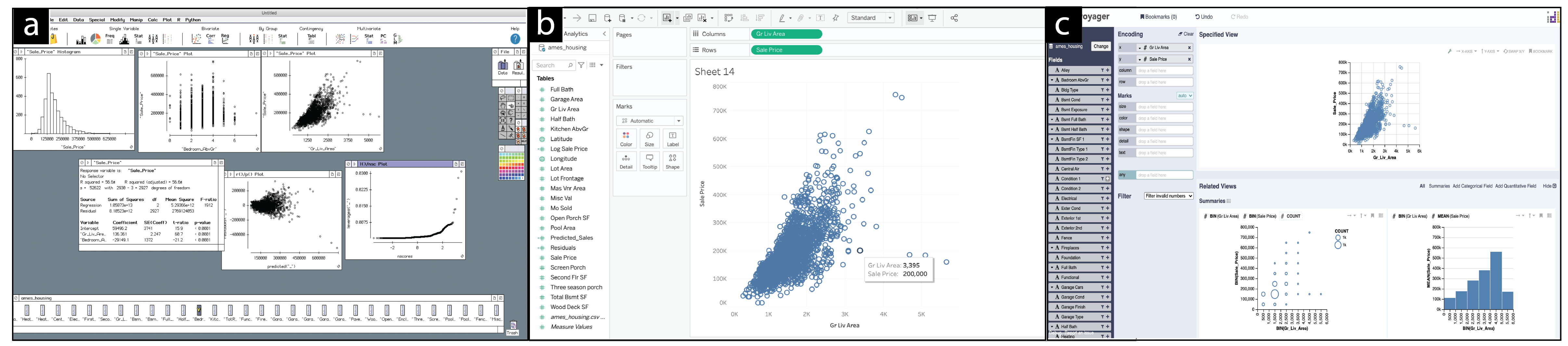}
  \caption{\it Three examples of GUI tools supporting exploratory visual analysis. a) DataDesk provides point-and-click visualization generation as well as easy access to statistical modeling functions like regression and associated diagnostic plots. b) In Tableau, dragging variables of interest to ``shelves'' above the plotting area results in a recommended visualization given principles of effective visualization design and a powerful table algebra~\citep{stolte2002polaris}. c) Voyager2 builds on the Tableau drag-and-drop approach by enabling easy exploration of many related views given selected data variables~\citep{wongsuphasawat2017voyager}. 
 }
  \label{fig:interfaces}
\end{figure*}

While notebooks created with RStudio, Jupyter, and similar packages are based in programming languages that offer considerable flexibility in terms of graphical and statistical functions, interactive graphical user interfaces for data analysis provide a more constrained environment, in which tabular data can be plotted and explored at the click of a button or dragging action of a variable. Among systems specializing in visualization interfaces to data analysis, Tableau Software~\citep{tableau} 
is perhaps best known for its focus on visualization-based data exploration and reporting. Tableau's interface is powered by combining principles from the grammar of graphics~\citep{wilkinson2012grammar} with a back-end table algebra that translates user interactions to visualization specifications associated with relational database queries~\citep{stolte2002polaris}. Other systems (Microsoft PowerBI~\citep{powerbi}, 
Oracle Analytics~\citep{oracle},
SAS JMP~\citep{sasjmp}, DataDesk~\citep{datadesk},
etc.) similarly advertise the power of visual analysis and offer their own versions of visualization-based front-ends for data exploration.  
These tools can vary in how much support they provide for different stages of analysis. A few systems, such as DataDesk (Figure~\ref{fig:interfaces}a) and JMP, provide graphical tools like brushing and linking as well as a suite of modeling tools to support canonical statistical models like regressions and hypothesis tests. Many other popular systems, such as Tableau Desktop (Figure~\ref{fig:interfaces}b) and Power BI, which are commonly adopted as visual analysis and reporting tools in applications like business intelligence, offer relatively little support for modeling and statistical testing.  These differences imply a question: What is the right amount of integration of statistical modeling functionality in a graphical user interface tool for exploratory analysis? 

If we look to research on state-of-the-art graphical user interface tools for exploratory and visual analysis, researchers often motivate their work in ways that imply that the value of the interface is to get out of the way of the data, so the human analyst can find the patterns or ``insights'' they hold. This view implies that advanced statistical modeling support is not a critical feature of a good visual analysis tool. Instead, tools are intended to create a responsive environment where queries are met at the ``speed of human thought''~\citep{heer2012interactive} and implement forms of ``behavior optimization''~\citep{rahman2020evaluating}, from visualization recommendations~\citep{vartak2015s,vartak2017towards,wongsuphasawat2015voyager,wongsuphasawat2015voyager} (Figure~\ref{fig:interfaces}c) to natural language interaction~\citep{gao2015datatone,setlur2016eviza,srinivasan2017orko} to big data optimizations~\citep{moritz2017trust}. These innovations aim to support more flexible inputs by which users can query and analyze data and to efficiently summarize data despite the scalability problems that arise as datasets grow larger. 

One possible presumption behind prioritizing data exposure in building these tools is that exploratory and confirmatory stages of an analysis workflow are easily distinguished. Some accounts of how knowledge is created during data analysis would seem to imply that so-called exploratory analysis is ``model free'' and consists of preparing and familiarizing oneself with data, searching for useful representations or transformations, and noting interesting observations~\cite{sacha2014knowledge}. Confirmatory analysis, on the other hand, involves verifying that data support a hypothesis~\citep{keim2008visual,pirolli2005sensemaking,sacha2014knowledge,thomascook}. 
Statisticians and others have long warned that a failure to distinguish exploratory and confirmatory stages can lead to ``naive empiricism run amok''~\citep{macdonald1983exploratory}, referring to pseudo-scientific use of data to confirm existing beliefs or identify patterns that do not betray underlying regularities in the target phenomena. 
Inappropriate overlap between exploratory and confirmatory analysis has even been proposed as a contributing factor to failed attempts to replicate what were believed to be high quality experiments in psychology and other fields, also known as  the ``replication crisis'' (e.g.,~\citep{nosek2018,wagenmakers2012}).
The dangers of too much overlap between EDA and CDA have more recently been the premise of work in computer science that pursues algorithms and interfaces for mitigating the symptoms of too much flexibility~\citep{pu2018garden,wall2017warning, zgraggen2018investigating,zhao2017controlling}, such as by tracking and adjusting for visual comparisons that an analyst makes~\citep{zgraggen2018investigating,zhao2017controlling}.

\begin{figure*}
\centering
 \includegraphics[width=\textwidth]{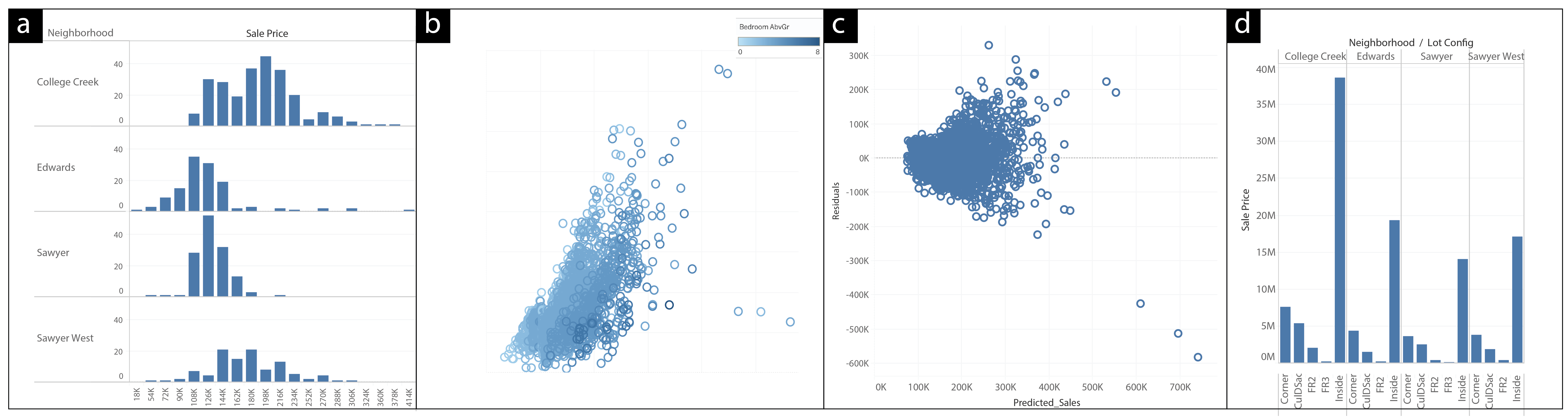}
  \caption{\it Plots of real estate data from Ames, Iowa~\citep{de2011ames}, \it created in Tableau Software~\citep{tableau}. (a) Trellis plot of  housing sale prices by neighborhood might invoke comparisons to a normal or log-normal distribution, and enables a visual check for a main effect of neighborhood. (b) Scatterplot of housing sale prices by square feet of above ground living area, with number of bedrooms above ground mapped to sequential color might invoke comparisons to a diagonal line representing a perfect positive correlation. (c) Residuals from multiple linear regression of sale price by above ground living area and bedrooms implied in plot b. (d) Trellis plot of sale price by lot configuration and neighborhood enables, among other effects, a visual check for an interaction between lot configuration and neighborhood.
 }
  \label{fig:four_ames}
\end{figure*}

In practice, however, it can be difficult to draw a clear line between exploratory and confirmatory data analysis. Model-driven inference plays a role even in canonically exploratory activities; after all, what is surprising is defined by the implicit or explicit model of our expectations. With the help of our visual system, we engage in processes comparable to fitting implicit models to data when we examine visualizations for distribution and trend, and we judge fit when we notice outliers and other deviations from symmetries inherent in graphical forms like histograms or scatterplots. We build up faceted displays like trellis plots to look for more complex effects and possible interactions in data. Conversely, we use graphs to assess residual deviation from models we explicitly specify and fit on data. Figure~\ref{fig:four_ames} depicts plots of real estate data that an agent might generate using Tableau in order to guide their strategy in choosing homes to represent, by getting a sense of distributional features and checking for a main effect of neighborhood in one part of a city (a), investigating the strength of relationships between living area, number of above-ground bedrooms, and sale price (b), explicitly checking residuals from a linear model that predicts sale price from living area and number of above-ground bedrooms (c), and checking for variance in the effect of lot configuration on sale price across neighborhoods (d). 


In this article we consider how assumptions about the analysis process and specifically the distinction between EDA and CDA may be reflected in interactive systems for exploratory visual analysis. 
We propose that designing software to strengthen, rather than separate, the links between purely exploratory and model-driven analysis can lead to better analysis. 
We argue that this bridging necessitates engaging with theories aimed at describing human statistical inference during graphical analysis. 
without an underlying theoretical basis to ground how exploratory activities feed the development of theories and models, computer scientists and statisticians can easily end up designing software that encourages only vague theories about how data were generated and conflicts with real world analysis stakes and goals. 
While our article is conceptual in nature, our arguments are backed by a growing amount of empirical research in the areas of interactive and exploratory analysis and uncertainty visualization.

The article is organized as follows: We first consider the origins of interactive data analysis, and how they might have led to a fixation in system design on exposure, the ``laying open of the data to display the unanticipated''~\citep{tukey1966data}. 
We describe how examples of negative implications of data exposure in recent research can be linked to an idea of ``rough CDA'' as a frequent activity in analysis.
We describe how conceiving of exploratory analysis activities as driven by model checks in a Bayesian framework provides a generalizable framework for developing interactive analysis tools to support multiple proposed stages of exploratory data analysis.
We note how this view compares to several other recent approaches to formalizing the role of statistical graphics in inference, including graphical inference as Bayesian cognition~\citep{kim2019,kim2020bayesian} and statistical hypothesis testing~\citep{buja2009,wickham2010}.
We discuss design implications of adopting a Bayesian model check formulation for interactive analysis software, and address related risks and open questions, including cognitive load implications and fundamental questions about what types of statistical problems visual analysis can address.  
We recommend ways in which interface features might better support analysts in specifying and testing implications of their implicit statistical models, from data diagnostics phases to rough confirmatory analysis. Finally, we discuss how attempting to fully automate a human-like analysis workflow might stimulate insights about how to improve interactive analysis interfaces.

\section{Background: Exploratory and interactive data analysis}
\subsection{Tukey on exploratory data analysis}
It may seem quite obvious that if you are doing data analysis, the interface you use should above all prioritize representation and easy access to the data. This way of thinking owes much of its motivation to the exploratory data analysis movement pioneered by John Tukey in the 1960’s. \citet{tukey1962future} popularized the idea of exploratory data analysis (EDA) as a natural complement to confirmatory data analysis (CDA), writing: ``The simple graph has brought more information to the data analyst's mind than any other device. It specializes in providing indications of unexpected phenomena.'' 

The proposal of EDA is memorable in part because he directly addressed a tension between the flexibility in thinking required to learn from one's data through construction of graphics and transformations and the supposed guarantees of confirmatory approaches. For instance, \citet{tukey1966data} wrote: ``Formal statistics has given almost no guidance to exposure; indeed, it is not clear how the informality and flexibility appropriate to the exploratory character of exposure can be fitted into any of the structures of formal statistics so far proposed'' and accused the formal inquiries into properties of confirmatory methods he saw around him as a means of ``legitimizing variation by confining it by assumption to random sampling'' and ``restoring the appearance of security by emphasizing narrowly optimized techniques and claiming to make statements with `known' probabilities of error.''  Tukey's classic text on EDA distinguishes it as a separate stage of analysis from CDA, and much of his work acknowledges a need to distinguish explicit confirmatory procedures to address implications of flexibility, stressing, for example, the importance of  treating as provisional identified patterns that had not been tested on different data through procedures like cross validation \citep{mosteller1977data}. 

It is not surprising then, that scholars have continued to stress a division between EDA and CDA, describing EDA as ``freewheeling search for structure'' \citep{buja2009} and repeating the analogy originally put forth by \citet{tukey1972exploratory} of a detective developing hunches while classically CDA activities like hypothesis testing can be likened to a jury deciding whether a defendant is guilty \citep{behrens1997principles,behrens1996data,buja2009,wickham2010}.

However, stressing a distinction betwen EDA and CDA can risk overlooking the strong sentiment in many of Tukey's writings of how exploratory analysis and model fitting go hand in hand. Some of the graphics he espoused can be interpreted in terms of a model; ``hanging rootograms,'' for example, would be difficult to motivate without reference to a Poisson count model. 
By attempting to fit models to data, one learns about what doesn't fit, a process that has been called model diagnostics \citep{buja2009} which enables that which didn't fit to be ``more effectively approached and structured because there has been some fit, even a poor one'' \citep{tukey1966data}. 

\subsection{Exploration versus confirmation in science reform}
The modern science reform literature also involves debates over the role and proper ``control'' of exploratory analysis in empirical scientific research. Many reformers have suggested that the lack of reproducibility of many high profile empirical results in psychology known as the replication crisis can be attributed to researchers failing to adequately separate exploratory from confirmatory stages of research. For example, \citet{wagenmakers2012} motivate an agenda for ``purely confirmatory research'' due to how exploratory analyses cause statistics to lose their guarantees. \citet{nosek2018} describe how overlooking the difference between EDA and CDA, such as by treating an exploratory study as though it were confirmatory after learning from the results, can ``lead to overconfidence in post hoc explanations (postdictions) and inflate the likelihood of believing that there is evidence for a finding when there is not'' (p.\ 2600, as cited in \cite{szollosi2019arrested}). Methods like preregistration \citep{nosek2019preregistration} in which a researcher declares their hypotheses and analysis plan in advance of data collection, have been popularized as solution.  

However, a growing body of work argues that attempts to strictly separate exploratory and confirmatory analysis are not well motivated logically or empirically. 
\citet{szollosi2019preregistration} argue that preregistration does not directly solve the problem of poor diagnosticity of statistical tests when exploratory findings are confirmed, since these depend critically on how well statistical models map to underlying theories, nor is there good reason to believe that it will encourage researchers to reflect more deeply on their theories, methods, and analyses. Others argue that some problems associated with a lack of distinction between EDA and CDA, like hypothesizing after results are known (HARKing), are not well evidenced to contribute to a lack of replicability \citep{rubin2017does} and can be helpful if done transparently \citep{hollenbeck2017harking}. \citet{devezer2020case} point out how the reform literature provides no unambiguous definitions for confirmatory versus exploratory.

Providing some support for a notion that EDA is ``model free,'' \citet{oberauer2019addressing} argue that in discovery-oriented research, theories do not strongly imply testable hypotheses. Instead, theories define a search space for effects that would support them, where failure to find effects does not invalidate theory. The question is not how the theory is wrong when effects aren't found, but why the data being assessed might not have been appropriate. Only in theory-testing research does the theory strongly imply a hypothesis and a lack of support for the hypothesis evidence against the theory. 
\citet{devezer2020case} describe how, in light of an alternative view that exploratory analysis often involves deliberate and systematic attempts at discovering generalizations (\cite{stebbins2001} as cited in \cite{devezer2020case}), exploratory analysis can be thought of as analogous to mapping unknown spaces until one is ``convinced that there is no element within the region being explored that remains undiscovered,'' whether these be theoretical spaces, model spaces, or concerned with experimentation. They relate hypothesis generation that occurs during exploratory analysis to abduction proper, in which scientists consider all of their knowledge about a phenomena with the aim of adding new insight or understanding, a process which is believed to be irreducible to formal statistical inference \citep{blokpoel2018deep,van2020theory}. 

Our view on philosophies of exploratory data analysis for designing interactive interfaces agrees with recent science reform discussions arguing that the relationship between exploratory and confirmatory activities is not as simple as proposals for clear separation between the phases imply. 
Like recent philosophical work in science reform, we acknowledge that there may not exist a normative model to encompass the diversity of activities associated with exploratory analysis.
We argue that attempts to formalize inference processes are nonetheless important for guiding interface design despite their imperfectness. This is because formalizations establish testable implications to drive knowledge gain about how EDA  occurs, while viewing EDA as atheoretical approach can restrict analysts from identifying connections between their graphical inferences and the models that would allow them to formalize them.  We cite empirical evidence suggesting that GUI EDA applications may encourage intuitive probabilistic inference that is not followed by confirmatory analysis~\citep{mccurdy2018,nguyen2020exploring,zgraggen2018investigating,zhao2017controlling}, motivating a need to better integrate support for activities associated with both EDA and CDA.

\subsection{Innovations in graphical user interfaces for analysis}

Modern interactive data analysis also owes much to developments in computer science, in the same way that earlier advances in statistical modeling by Laplace, Gauss, and so forth accompanied progress in mathematics. As Tukey began writing about exploratory data analysis, computer scientists such as Engelbart, Kay, Sutherland, and others made pioneering efforts in the development of software interfaces for ``intelligence augmentation.'' As promoted by \citet{engelbart1963}, intelligence augmentation is associated with ``increasing the capability of a man to approach a complex problem situation, to gain comprehension to suit his particular needs, and to derive solutions to problems.'' Increased capability could come as greater efficiency (perhaps framed as ``more rapid comprehension'' or ``speedier solutions'' as well as improved perception of possible solutions to problems that before seemed unsolvable. The broad framing of IA by these early pioneers outlined a vision for transforming interactions with computers in which graphical user interfaces for data analysis were a natural step.

Tukey’s and colleagues' system PRIM-9 augmented the capabilities of a human by enabling perception of higher (2+) dimensional data \citep{fisherkeller1988prim}. A user could ``dissect'' multivariate data through point cloud rotation, use masking to select subregions of a space, and isolate particular subsamples. Because an analyst will rarely be able to specify the ``optimal'' projection, finding an appropriate one requires moving about in a multidimensional space, which PRIM-9 enabled through controlled continuous rotation \citep{friedman2002}. Tukey’s work on PRIM-9 led to further developments through projection pursuit, the incorporation of automation into interactive visualization by optimizing a projection index to detect interesting directions of study \citep{friedman1974projection}. 

In the decades that followed, other statisticians made graphics contributions. Asimov~(\citeyear{asimov1985grand}) introduced the grand tour, which used animation to stitch together projections on high dimensional data for visual analysis in a seemingly continuous way. Projection pursuit guided tour combined both methods for better results when identifying low-dimensional structures in sparse high dimensional data \citep{cook1995grand}. \citet{becker1987brushing} explored brushing as a way to interactively select data in a visualization one is analyzing, in order to see the same data in other linked views, such as when viewing a scatterplot matrix. XGobi \citep{swayne1998xgobi}, followed by GGobi~(\cite{swayne2003ggobi}), made these state-of-the-art dynamic statistical graphic methods available in a single environment. The ``scagnostics'' (scatterplot diagnostics) of \citet{wilkinson2005graph,wilkinson2006high} explored a graph-theoretic set of measures for grouping bivariate scatterplots of high dimensional data, and the grammar of graphics \citep{wilkinson2012grammar} provided a formal description of statistical graphics.

Computer scientists also began to take more interest in the new interactive capabilities for data analysis afforded by more powerful computation. \citet{shneiderman1974computer,shneiderman1982future} coined the term ``direct manipulation'' in the early 1980s to refer to systems in which objects of interest such as data points were continuously represented and could be acted on through physical manipulation or button presses. In contrast to the inflexible and hard to learn syntax of conventional query languages, direct manipulation was easy and produced immediately visible and reversible results \citep{hutchins1985direct}. One could call direct manipulation interfaces for data analysis an early step towards ``democratizing data analysis,'' as these tools reduced the amount of specialized knowledge required to interact with data; one no longer needed to memorize rigid syntax, for example.

The late 1980s saw the emergence of visualization as a subfield of computer science \citep{mccormick1987visualization}, focused on amplifying cognition through visual methods drawn from computer graphics, vision, signal processing, human computer interaction, and others, and addressing domain applications like medical imaging, planetary sciences, and molecular modeling. Information visualization, which is closer to our focus here, concerns visualizing abstract data for which spatial mappings can be chosen more arbitrarily (e.g., statistical graphics) and was distinguished in the 1990s \citep{card1999readings} drawing in cognitive scientists and psychologists, statisticians, and cartographers. While many early advances sought to enhance data analysis among experts, the last few decades of research in the field has seen a surge of interest in making visual data analysis accessible to more novice users. Today, widely used systems like Tableau Software employ innovations that grew out of visualization research, by encoding state-of-the-art knowledge on effective visualization \citep{mackinlay1986automating} and reducing the efforts required to manually specify views through drag-and-drop interfaces like Tableau’s shelf model \citep{tableau,stolte2002polaris} or button-driven chart type  transformations \citep{mackinlay2007}, which interpret these user interactions as database queries.

More recently, recommender systems have become an active area of research in visualization \citep{vartak2017towards}. Recommenders aim to be even more hands off than popular visualization tools like Tableau or PowerBI by suggesting data wrangling operations \citep{kandel2011wrangler}, views to analysts based on perceptual properties \citep{wongsuphasawat2015voyager,wongsuphasawat2017voyager}, statistical analyses \citep{demiralp2017foresight,key2012vizdeck,vartak2015s} or contextual or behavioral properties \citep{bromley2014dive,lin2020dziban,key2012vizdeck,gotz2009behavior}, requiring minimal to no input from the user after a dataset has been loaded. Other tools literally make analysis hands-off or at least ``mouse off'' by supporting new input modalities like natural language \citep{gao2015datatone,setlur2016eviza,srinivasan2017orko} or touch \citep{powerbi,vizable}. 
Such forms of ``behavior optimization'' comprise the state of the art in interactive data analysis system design \citep{rahman2020evaluating}.

Research in visual analytics has evolved in tandem with that in interactive visualization, being differentiable mainly in its focus on integrating visualization-based and automated data analysis methods and on large datasets that motivate such automation~(\cite{keim2008visual,thomascook}). Relevant to our interests are attempts to conceptualize the visual analytics process as a model of knowledge generation \citep{andrienko2018viewing,keim2008visual,pirolli2005sensemaking,sacha2014knowledge,wang2009defining}. Researchers have commented on modeling and uncertainty as implicit in exploratory and interactive analysis \citep{andrienko2018viewing,sacha2014knowledge}. 


Similarly, the rise of ``Big Data'' as a fascination and challenge faced  by industry has also driven increased interest in interactive analytics in database research, referring to approaches for optimizing query results for real time analysis by a human. These applications bring their own challenges \citep{andrienko2020big,fisher2012interactions}, such as minimizing latency while retaining acceptable accuracy. User interfaces have not always been central to these efforts, but how to deliver visualizations and interactions in these paradigms is gaining interest \citep{alabi2016pfunk,fekete2016progressive,kim2015rapid,moritz2017trust,park2016visualization}.

\section{Exploratory analysis or rough confirmatory analysis?}
That exploratory visual analysis systems support a diverse range of activities--from data diagnostics, to characterizing distributions and relationships, to looking to support a hypothesis---is acknowledged in the literature on interactive visual analysis (see \cite{battle2019characterizing} for a review). Descriptive accounts also tend to acknowledge that it often alternates between open-ended tasks (e.g., flipping through filters looking for something interesting to explore a space of theories or models (a.k.a, abduction proper; \cite{devezer2020case,oberauer2019addressing}) and more focused exploration (e.g., trying to formulate and validate a hypothesis). However, recent analogizing of exploratory visual analysis to a multiple comparisons problem by computer scientists \citep{pu2018garden,zgraggen2018investigating,zhao2017controlling} emphasize what \citet{tukey1972data} referred to as ``rough confirmatory analysis.''

\citet{tukey1972data} characterized analysis as beginning with an initial exploratory phase in which the analyst doesn't consider probability, followed by an intermediate probabilistic stage in which the analyst attempts to answer the question, ``With what accuracy are the appearances already found to be believed?'', followed by confirmatory testing. In the intermediate, rough confirmatory stage, an analyst seeks a coarse set of possible answers to the question of how accurate the apparent patterns are. He described how the appearances could be ``so poorly defined that they can be forgotten,'' or marginal (such that "crude analysis might not suffice and a more careful analysis is called for''), or well determined such that ``we may, but more often do not, have grounds for a more careful analysis.'' Hence, visual analysis is framed as playing a classification role in helping an analyst distinguish between signals so obvious that statistical modeling is not needed, to those where noise and confounding might be so great that confirming any perceived patterns is hopeless. Tukey stressed multiplicity as a key issue in this second stage  \citep{tukey1972exploratory,tukey1969,tukey1972data}, including ``How many things might have been looked at? How many had a real chance to be looked at? How should the multiplicity decided upon, in answer to these questions, affect the resulting confidence sets and significance levels?'' 

\subsection{Empirical critiques of ignoring inference in exploratory visual analysis}
An emerging line of critically-themed research in interactive visualization and analysis attempts to problematize a model-agnostic approach to designing software for visual analysis, implying that users of exploratory visual analysis tools frequently engage in rough confirmatory analysis. 
Much of this work remains speculative, suggesting by way of examples how different types of cognitive biases may arise in interactive analysis \citep{dimara2018task,wall2017warning}. However, a growing number of empirical studies are being used to argue about potential threats to valid inference from flexibility or design decisions in exploratory visual analysis.


For example, one recent line of research argues that by enabling the user to query more and faster, modern interactive systems for data analysis are particularly likely to result in a multiple comparisons problem~\citep{pu2018garden,zgraggen2018investigating,zhao2017controlling}. 
When using standard approaches to null hypothesis significance testing (NHST), a multiple comparisons problem arises because NHST admits a certain percentage of false positives by definition. Hence the more tests one does, the more false positive conclusions one might expect to arrive at.
An implication made in some recent interactive analysis research is that if visual comparisons are analogous to significance testing, where a $p$-value is used to judge whether an effect can be ruled unlikely to be due to chance, as some statisticians have proposed~\citep{buja1999inference,buja2009,wickham2010}, then those developing interactive analysis system should introduce measures to control the potential to produce false discoveries.

\citet{zhao2017controlling} described how most people in a sample they studied who were looking at histograms of Census data in an analysis task treated patterns they saw as if they were reliable (``significant'') and didn't consider how the number of comparisons they did inflated their chance of finding something interesting. This might suggest that visual analysts stick to observing only very large patterns where follow up analysis is unnecessary. However, \citet{zgraggen2018investigating} estimated the severity of the multiple comparisons problem among 28 moderately experienced analysts, who used an interactive visual analysis tool to identify any reliable observations or recommendations as they assessed data samples they generated from a known ground truth population. The authors tracked each analyst's total number of visual comparisons using a combination of experimenter questioning and eye tracking, and used statistical tests against the ground truth to determine the accuracy of each type of observation they saw (e.g., a comparison between two groups, a statement about the shape of a distribution, etc.). This led to an estimate that over 60\% of the analysts' conclusions were spurious. 

Using a similar prompt asking analysts to report generalizations that could be made from exploratory visualizations, \citet{nguyen2020exploring} investigated how plotting defaults in interactive visualization and business intelligence tools like Tableau Software may affect novice analysts, who tend to be least likely to know when or how to change from a default setting in software \citep{shah2006policy}. In two online experiments, they showed participants data samples using either disaggregated views, mean aggregated views, or disaggregated views with an overlaid mark showing the mean and asked what they might conclude, if anything, about a population. They found that those who used disaggregated views were less than one-fifth as likely to talk about effects without mentioning how big they are (e.g., ``There's no difference in sales between campaigns,'' ``Visitors from the midwest bought more''). They reported lower confidence values by an average of 6 points on a 100 point scale, and showed more sensitivity in terms of how many conclusions they drew to whether they were looking at 50 records or 1000 records.

\begin{figure*}[t]
    \centering
    \includegraphics[width=\textwidth]{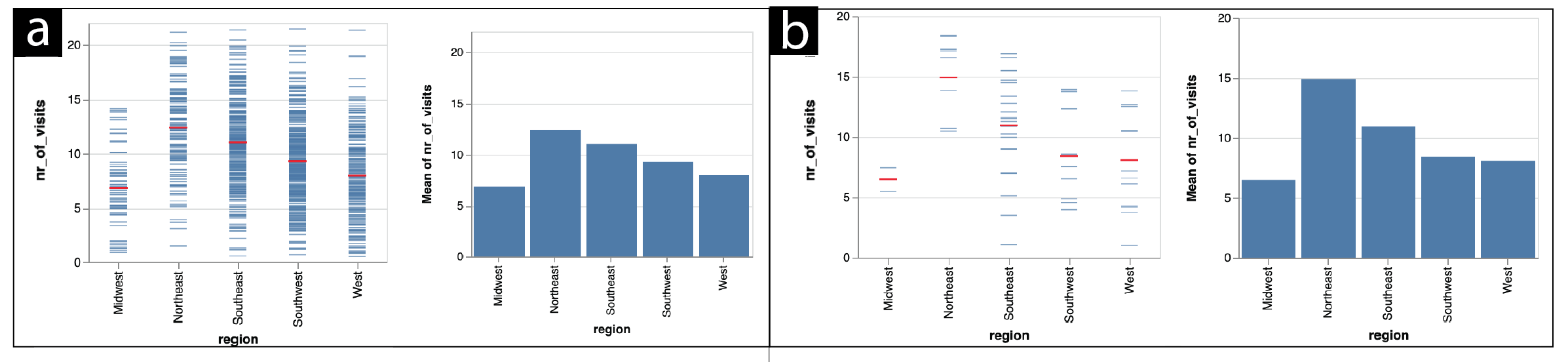}
    \caption{\it Plots showing the number of clicks on online advertisements by the region associated with the website visitor's network domain, such as a web marketing analysis might examine to look for regional patterns. Plots are generated from a dataset of size 1000 (a) versus 50 (b), and visualized using two possible default plotting approaches an exploratory visual analysis interface might adopt: showing disagreggated data by default but annotating the mean (left side of each panel), or using only mean aggregation (right side of each panel). \citet{nguyen2020exploring} find that the aggregation default affects how sensitive novice data analysts are to effect size.}
  \label{fig:agg_vis}
\end{figure*}

These recent empirical studies have taken issue with the ambiguity of the concept of an ``insight,'' which is commonly used to characterize conclusions drawn from an interactive analysis session \citep{rahman2020evaluating}. This term has been defined in various ways, with one common definition being a ``complex, deep, qualitative, unexpected, and relevant revelation'' \citep{north2006toward}. While insights are often framed as being closer to confirmatory processes---\citet{sacha2014knowledge}, for instance, distinguish between exploratory ``findings'' and more formalized verification loops that involve ``hypothesis'' and ``insight''---rarely are degrees of belief in an insight discussed or elicited. 
A recent empirical study on professional analysts' naturalistic insight generation with visualization tools found that only a handful mentioned that identifying an insight involves consider how confident one can be in it~\cite{law2020data}, echoing the insensitivity to probability in analysis conclusions described by the aforementioned studies. Again, this suggests either that users of interactive visual analysis tools comment only on very large, obvious patterns, or that they are sometimes drawing conclusions from visual evidence alone, in cases where follow-up analysis would be well-motivated. The growing empirical evidence seems to better support the latter interpretation.

Several recent critiques in the interactive visualization literature also point to the absence of attempts to elicit or formalize the role of prior knowledge in interactive analysis studies or systems \citep{kim2019,koonchanok2021data}.
Though research in visual analytics implies that prior knowledge plays a role in what one considers a finding or insight \citep{federico2017role,lammarsch2011towards,mccurdy2018}, few attempts have been made to integrate prior knowledge into visual analysis beyond allowing analysts to link text notes to views.
For example, \citet{mccurdy2018} conclude from an empirical study of visual analysis by global health experts that the experts often mentally adjust the data they see to account for known ``implicit'' error, but the authors imply that such knowledge cannot be integrated directly with data representations. Cognitive psychologists have studied how experts' prior knowledge and reasoning strategies lead them to interact with visualizations differently than novices (see, e.g., \cite{hegarty2004diagrams,trafton2000turning}), yet the findings of this literature have not necessarily influenced the design of exploratory analysis tools.  
One recent exception is a Wizard of Oz study by \citet{choi2019concept}, which explores how well users of an exploratory visualization tool can articulate conceptual and model-based expectations they bring to data based on their prior knowledge, finding that they frequently used visualizations to validate their expectations.

Limitations of common evaluation methods for interactive visualizations and analysis tools have been another point of critique. Researchers evaluating interactive visual analysis tools often use lower time spent on a task and lower error in responses as desirable criteria, along with reported satisfaction through qualitative user feedback (see \cite{rahman2020evaluating} for a review), similar to the evaluation of interactive visualization more broadly, where reliance on accuracy reading data and response time have motivated a long running workshop entitled ``Beyond Time and Error'' \citep{beliv}. These measures are common even when the goal of an interactive visualization is framed as supporting reasoning under uncertainty (see \cite{hullman2018pursuit} for a review), suggesting that researchers may not know how to define measures that would better capture inference or decision quality. 

These and other critiques imply that inference \textit{is} an important goal in visual analysis. While this might appear obvious to many readers, the idea that, as \citet{tukey1990data} described, phenomena---referring to potentially interesting things that we can describe in non-numerical terms---are what we typically want to learn about when we deal with data, has not been emphasized as much as the idea of immediate support for pattern finding. Such critiques motivate our proposal that research on supporting exploratory visual analysis should embrace theories of graphical inference. In the following section we propose an alternative understanding of exploratory visual analysis as guided by model checks, and describe possible formalizations of this theory.  

\section{A Bayesian theory of inference for interactive analysis}
\label{sec:theories}

The microcosm of activities that comprise interactive analysis---from data diagnostics to theory exploration to rough to proper confirmatory analysis---may help explain why design philosophies behind the development of interactive analysis interfaces are hard to identify and at times seem in conflict. Despite the diversity of activities that occur in data analysis, we propose that research aimed at developing better interfaces for exploratory analysis would benefit from a more formal approach to defining the mechanism behind human graphical inference during interactive analysis. 

We motivate the need for a theoretical model as follows. Even if some activities fall outside of the predictions of any specific model, without an underlying theoretical framework to guide the design of tools, we are hard pressed to identify where our expectations have been proven wrong and can easily end up with the sort of piece-meal and mostly conceptual theories that dominate much of the literature on interactive analysis.
This lack of formalization makes it difficult to falsify or derive clear design implications from theoretical work. 
For example, some work suggests that visual analysis is a process of fitting intuitive models \citep{andrienko2018viewing,choi2019concept} or sensemaking under different forms of uncertainty \citep{sacha2014knowledge}. However, ambiguity in the underlying assumptions about the structure of an intuitive model and how it may evolve given a sequence of analysis operations render these conceptions hard to falsify.


An important clarification is that the value of proposing formal theories of graphical inference does not depend on those theories or the goals they imply being completely accurate. As researchers we may never be able to define what it means for an EDA process to be ``optimal'' or to perfectly predict human graphical inference in a given situation. However, part of our goal toward improving interactive analysis interfaces should be to propose and evaluate theories of human inference that are applicable to many instances, and may provide a standard for instances that exhibit clear deviation from principles of statistical inference.

What might a formal theory to describe how an analyst responds to data during interactive analysis look like? If, as the various literature seems to suggest, it is difficult in practice to distinguish EDA from CDA beyond the fact that CDA is a ``final'' step of confirming one's inferences about real world phenomena, then the theory of statistical inference should provide a useful prescriptive grounding for such a formalization.
We motivate an understanding of interactive visual analysis as a process of implicit model checking, then formalize this idea in a Bayesian statistical framework. We discuss this understanding in comparison to related proposed theories of visual analysis.





\subsection{Implicit model checking in interactions with data}
At a high level, if EDA is understood to be discovery of the unexpected as is generally assumed, then this is defined relative to the expected. We note two practical implications of this duality:
\begin{enumerate}
    \item Any exploratory graph should be interpretable as a model check, a comparison to ``the expected.'' This implies that when constructing such graphs we should be able to figure out what is the model being used as a basis of comparison. Sometimes, as with a residual plot (Figure~\ref{fig:four_ames}c), this comparison is obvious; in fact many exploratory graphics themselves get their meaning from implicit reference distributions, from histograms inviting comparisons to bell curves when they are remotely symmetric in appearance (Figure \ref{fig:prior_predictive}) to cumulative distribution plots inviting comparisons to the diagonal.  
    Other times we can gain insight by carefully considering what sort of model is being implicitly checked by a graph.  For example, a trellis plot of histograms of house sale price might be used by a real estate agent in Ames, Iowa to check for a main effect of neighborhood in western Ames (Figure~\ref{fig:four_ames}a). A trellis plot showing distributions of hours of sleep for sleep tracker users who report female versus male as their sex and who have and have not previously used sleep trackers (Figure~\ref{fig:post_predictive}a),  which might be used by an analyst at a sleep tracking company, can be interpreted as a check of, or exploration of discrepancies from, a more complex linear model that predicts ${\rm hoursSleep}$ using the three other plotted variables (e.g., ${\rm hoursSleep} \sim \alpha + \beta_{s}*{\rm male} + \beta_{t}*{\rm sleepTracker} + \beta_{f}*{\rm fitnessLevel}$).
    \item Exploratory analysis can be made more effective by comparing to more sophisticated models. EDA is often thought of as an alternative to model-based statistical analysis, but once we think of graphs as comparisons to models, it makes sense that the amount we've learned increases with the complexity of the model being compared to. Effective graphics create visual structures that enable model inspection by foregrounding comparisons of interest in ways that exploit the abilities of the human visual system~\citep{bertin}, such as to detect deviations from symmetry. Graphics are iterated on during exploratory data analysis to refine the visual comparison or increase the complexity of the  model, such as by adding additional variables to trellis plots, or calculating derived fields to isolate effects while still relying on position encodings. 
\end{enumerate}

There is a corresponding argument in classical hypothesis testing or confirmatory data analysis, that more is learned from rejection of a complex model than from rejection of a trivial null model such as a hypothesis that all effects are exactly zero. In some ways, EDA is like an omnibus test in that we are open to all sorts of violations of the model, but with the difference that in exploratory analysis we are interested not so much in rejection as in the particularities of the discrepancies between model and data: rather than tailoring tests to particular alternatives, we rely on human pattern-finding abilities to motivate the development of future hypotheses. For example, in examining a trellis plot like Figure~\ref{fig:post_predictive}a, an analyst might implicitly compare the rows of the trellis plot, and separately columns, to check for main effects of prior use of a sleep tracker and sex. 
They might scan for any panels that seem to deviate from the others to check if there appear to be any interactions between prior use of a sleep tracker, sex, and fitness level. Without necessarily realizing it, they might conduct a sort of ``mental cross validation,'' fitting a linear model to subsets of the data and then comparing to the left over panel each time. As a result, they might observe what they think is a slightly different effect of higher fitness level for males who previously used sleep trackers, and might make further graphical optimizations to assess this observation, such as adding trend lines (Figure~\ref{fig:post_predictive}b). 


To summarize this view, rather than assuming that analysts using interactive analysis software look for patterns only in a non-probabilistic mode, we instead conceive of them as developing and updating ``pseudo-statistical'' models that help them make inferences about real-world phenomena. By real world phenomena, we mean a referent for an observation made in data analysis that exists outside the numbers or strings that comprise the dataset. This might be a measurement process, as in identifying errors in data collection, or a data generation process, as in trying to ascertain explanations of variability or skew. These phenomena might be evaluated in a past, present, or future tense.

In contrast to statistical models an analyst might explicitly specify and test, we call these models pseudo-statistical because while they may be approximated statistically, they may deviate from what is generally defined as rational inference. For instance, they may be mentally represented in ways that deviate from a proper statistical model (e.g., at times neglecting probability information so as to explore a space of possible theories or explanations for data), and they may not be updated as predicted by a standard Bayesian model of belief updating. 
Importantly, because of the potential differences between the predictions of an analyst's pseudo-statistical model
and a corresponding proper statistical model fit to the observed data and available prior knowledge, we argue that GUI exploratory data analysis tools should support explicit model checks on the part of the user to aid them in adjusting their expectations. 
Below we make this proposal more concrete by formalizing it in a Bayesian framework, than then talk about its implications for the design of software.

\subsection{A Bayesian formulation of graphical inference as model checking}
Our proposal above is informed by a formulation of graphical inference and exploratory analysis as analogous to a ''model check'': a comparison of data to replicated data under a model, previously proposed by \citet{gelman2003,gelman2004exploratory}. 
Put simply, the model checking formulation says that in viewing graphics, the user imagines data produced by a process that seems reasonable to them, and compares these imagined data to the observed data plotted in the graph.

More formally, assuming a parameter(s) of interest $\theta$, the model checking formulation expands the notion of a posterior distribution in Bayesian inference from $p(y|\theta)p(\theta)$ to $p(y|\theta)p(\theta)p(y^{rep}|\theta)$. In the formulation of \citet{gelman1996posterior}, $y^{rep}$ is a replicated dataset of the same size and shape as the observed dataset $y$, but produced by a hypothesized model that accounts for what is known about $\theta$. 
All model checks, whether exploratory (driven by graphical comparisons) or confirmatory (driven by $p$-values), represent comparisons between $y$ and $y^{rep}$. Visualizations, whether real or imagined, can be thought of as visual test statistics ($T(y)$ and $T(y^{rep})$); in other words they play the role of summaries that capture the amount of signal in the data (observed or imagined). 

\begin{figure*}
\centering
 \includegraphics[width=\textwidth]{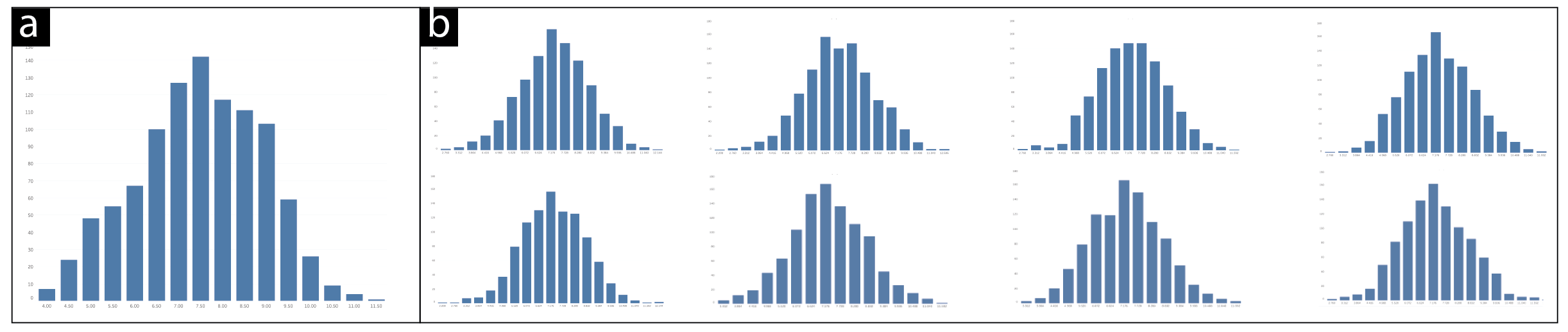}
\caption{\it An observed distribution of hours slept per night among 1000 simulated users of a sleep tracking app (left), compared to a lineup~\citep{buja2009} comprised of eight samples from a Gaussian distribution with the same location and scale as observed data. The skewed observed distribution deviates from expectation in that slightly fewer people than expected sleep for very long periods, and more people sleep between eight and nine hours.}
  \label{fig:skewed_hours}
\end{figure*}

To make this concrete, consider an analyst doing initial checks of distributions after loading data in a tool that offers interactive data transformation and visualization. They plot multiple quantitative variables of interest to histograms, and then inspect each to judge distribution. If the data appear even remotely symmetric, the analyst might naturally try to assess the degree of symmetry, implicitly comparing what they see to imagined normally distributed data from censored or non-censored distributions centered on the location they perceive in the plotted data. They might naturally attend to data that deviate from their expectations of tail behavior for the implicit distributions, or alternate between comparisons to different implicit reference distributions (e.g., unimodal versus mixture) to judge distribution shape. For example, an analyst working for a sleep tracker company might notice that relative to a Gaussian distribution with similar location and variance (represented by draws shown in Figure~\ref{fig:skewed_hours}a), the distribution of hours slept by 1000 sleep tracking app users shows fewer people sleeping more hours per night (Figure~\ref{fig:skewed_hours}a). The visual ``test statistics'' that the analyst perceives might be subjected to a discrepancy function, producing something like an implicit $p$-value to be judged against the analyst's internal criteria for when ``enough'' evidence exists for a claim~\citep{buja2009}.
Depending on the outcome, these model checks might be followed by the analyst seeking more information about the data collection process to determine the cause of perceived errors, by the use of statistical summaries or diagnostic tests of shape if the analyst plans to do confirmatory testing down the road, or simply by moving on to bivariate comparisons with more assurance that they understand outliers, skew, or other properties of the variables.

In a Bayesian framework, the hypothesized model that produces the data in these imagined histograms ($y^{rep}$) is the posterior predictive distribution $p(y^{rep}|y)$. This distribution can be viewed of a transformation of the posterior distribution $p(\theta|y)$ from parameter space (i.e., in terms of $\theta$) to data space (i.e., in terms of the underlying measurements). In a Bayesian statistical paradigm, the posterior distribution $p(\theta|y)$ is produced by updating a prior distribution $p(\theta)$ by applying Bayes' rule to a distribution $p(y|\theta)$ that captures the likelihood of different values of $\theta$.
The posterior predictive distribution $p(\theta|y) \propto p(y|\theta)p(\theta)$ is then calculated by marginalizing the distribution of $y_{rep}$, which we can think of as a newly drawn version of $y$ given $\theta$, over the posterior distribution of $\theta$ given $y$. Reflecting on this machinery, the Bayesian model check analogy suggests that an analyst's judgments about patterns in data are influenced by perceived properties of the observed data (likelihood), and by extension the set of visual encodings through which the observed data is perceived. They can also be influenced by the prior knowledge of the analyst, which can play a regularizing role by shifting observed differences between groups closer to one another, or the implicit posterior predictive distribution away from the location or scale of distributions inferred from the observed data. Finally, they are undoubtedly influenced by the analyst's statistical experience, since the space of possible models they perceive will depend on their knowledge.

\begin{figure*}
\centering
 \includegraphics[width=\textwidth]{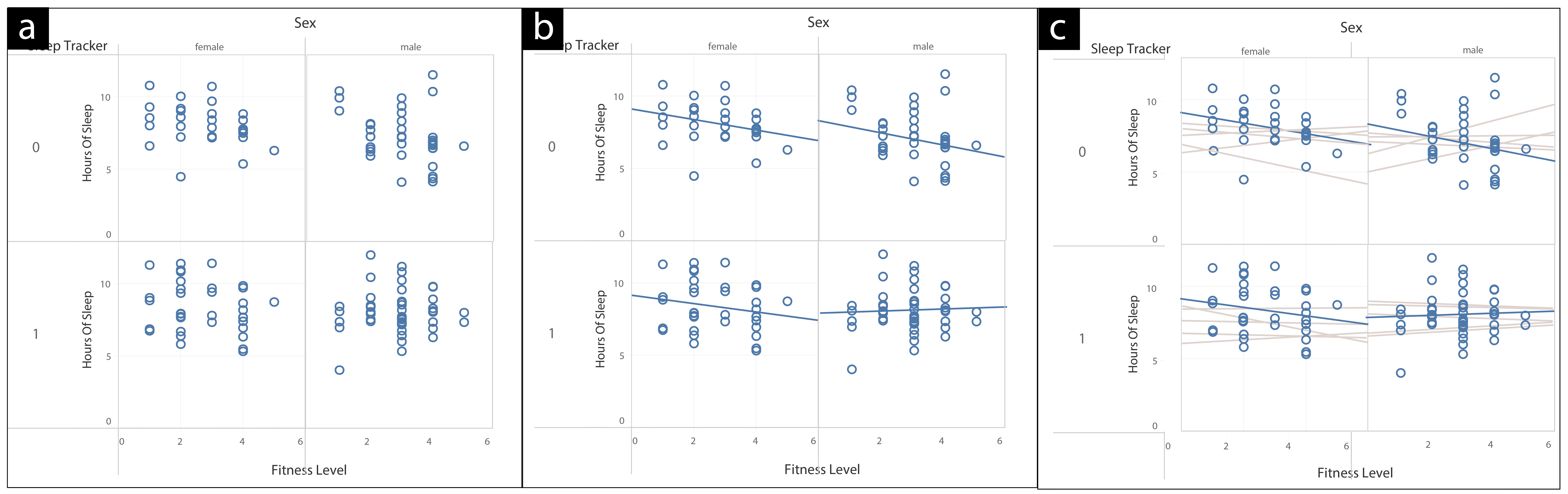}
 \caption{\it An analyst at a sleep tracker company might use a system like Tableau to create a trellis plot like that shown in panel a, observing what they perceive as a directionally-different effect of higher fitness level for males who previously used sleep trackers (shown more clearly in panel b by adding lines of best fit to each pane). Upon seeing random draws (panel c), which might be shown a few at a time via animation, or in a static ensemble or interval plot, the analyst might correct any gaps in their understanding of a reasonable data generating process.  }
  \label{fig:post_predictive}
\end{figure*}

Consider again the trellis plot in Figure \ref{fig:post_predictive}a.
As in most visualizations, the plot is intended for estimating more than one parameter, so $\theta$ is a vector, which might include slopes and intercepts for each combination of sex and sleep tracker, as well as more specific comparisons across fitness levels or hours of sleep within or across particular views.
The analyst might perceive slightly different relationships (e.g., different slope directions) between hours of sleep and fitness level for females versus males. 
In eyeballing the plot to estimate intercepts, they might consider prior knowledge they have about the average difference in hours of sleep between males and females informed by research on sleep trends (e.g., \cite{burgard2013gender}). The analyst's perceptions of trend in light of the available information could be compared to the predictions of a maximal (i.e., including all interactions) Bayesian regression model that accounts for this prior knowledge and places weakly informative priors on other variables. A few random draws from such a model are plotted in Figure~\ref{fig:post_predictive}c, and from even this small set it is evident that any small difference an analyst might perceive in slope is likely unreliable in light of the prior and high variance of the observed data.
. 

\begin{figure*}
\centering
 \includegraphics[width=\textwidth]{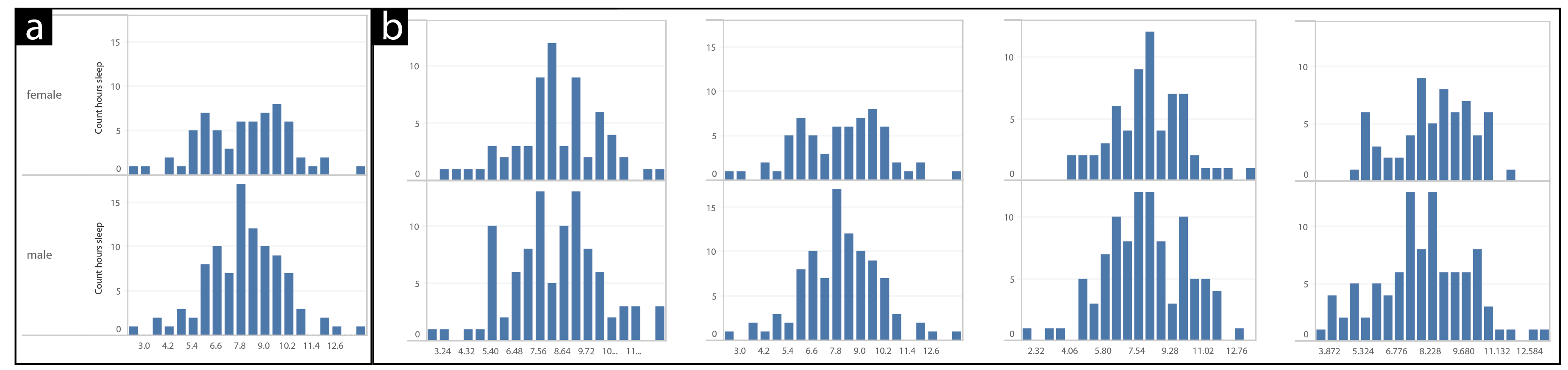}
 \caption{\it An analyst at a sleep tracker company might examine a trellis plot of histograms of observed hours of sleep (a) from 150 sleep tracker users of female (top) and male (bottom) sex compared to simulated data generated from a multivariate prior distribution based on a larger sleep survey~\cite{burgard2013gender} (b) to check whether their sample has any notable dissimilarities.}
  \label{fig:prior_predictive}
\end{figure*}

In a Bayesian statistical workflow, visualization is also used to reason about the appropriateness of the prior, and to compare its predictions to the observed data~\citep{gabry2019,gelman2020bayesian}.
For example, an analyst might examine the difference in the observed distributions of hours of sleep for female and male sex (Figure~\ref{fig:prior_predictive}) against draws from a prior predictive based on the prior research~\citep{burgard2013gender}.

\subsection{Relationship to other models of graphical inference}

\subsubsection*{Graphical inference as Bayesian cognition} 
In cognitive science, Bayesian models of cognition \citep{griffiths2008, griffiths2012} have gained traction for modeling various forms of human cognition, including object perception \citep{kersten2003}, causal reasoning \citep{steyvers2003}, and knowledge generalization \citep{tenenbaum2006}. These models assume individual cognition relies on Bayesian inference: an individual's implicit beliefs about the world are captured by a prior; when exposed to new information they update their prior according to Bayes' rule, arriving at posterior beliefs. Recent work applies Bayesian models of cognition to how people draw inferences when shown visualized data, either eliciting their prior beliefs about a parameter (e.g., \cite{karduni2020,kim2019,kim2020bayesian}) or endowing priors, showing them new data, and then eliciting their posterior beliefs to compare to normative Bayesian posterior beliefs from one or more models reflecting different ways that Bayesian updating could occur.

Bayesian cognition has been applied to visualization-based inference in a normative sense, where a Bayesian model is used to define ``good belief updating'' as a standard for comparing to or guiding people's belief updates from data (e.g., to evaluate different representations of uncertainty in data \citep{kim2019} or guide belief updates \citep{kim2020bayesian}). It has also been used in a more descriptive sense, in which observations of people's belief updates are analyzed to gain insight into how human inference deviates \citep{karduni2020,kim2019}, ideally approached using principled tools for model evaluation and model selection \citep{tauber2017bayesian}. Toward both normative and descriptive applications, the mathematical basis of Bayesian inference has been used to calculate measures of graphical inference like perceived sample size, the size of the equivalent random sample that a Bayesian would have needed to see to arrive at the posterior beliefs expressed by a user \citep{kim2019}. Toward more descriptive ends, a researcher might attempt to model sources of deviation from normative updating based on factors other than the statistical informativeness of the data. For example, hierarchical models in which hyperpriors describe the bias a person expects from a given information source can be used to  reflect on the forms and strength of distrust in data as a reason for deviation in some settings.  Integrating the predictions of perceptual models like implicit logarithmic perception \citep{Gonzalez1999,Hollands2000,Stevens1957,zhang2012} can help researchers separate cognitive and perceptual factors. 


How is a Bayesian cognition framework as applied to interactive visualization related to the Bayesian model check framework described above? Both theoretical frameworks rely on the generalizability of a Bayesian modeling framework for describing human inference. Both can be used descriptively or normatively. 
In many ways, their normative versions are complementary when considering applications to interactive analysis and visualization: Bayesian cognition emphasizes trying to achieve more rational updating in the context of a predefined model, while the model check formulation enables us to study what implicit model structures analysts assume when using graphs to reason about observed data in light of possible prior knowledge. 
Approaches informed by Bayesian cognition should help reduce the gap between analysts' expectations based on implicit models and reference distributions from the standpoint of information accumulation and statistical learning. 

What does it look like to use these complementary theoretical frameworks to improve people's behavior? Beyond the utility of these frameworks for studying and characterizing behavior, can they also be incorporated into systems to improve behavior on the fly? 
In particular, the common tendency people show toward under-updating as sample size grows, and over-updating as it shrinks (as described by a model of non-belief in the law of large numbers; \cite{benjamin2016}) implies testable implications for analysis software. 
Specifically, conservatism in belief updating deriving from a bias like non-belief in the law of large numbers would suggest showing more frequent, smaller samples.
Applied to a progressive computation or approximate query processing setting in which analysts are shown visualizations of partial query results on very large data, developers of interfaces might think twice about design strategies that provide an initial partial result then only alert the user to check results when the queries are finished. 
Applied more broadly to systems for exploratory visual analysis, conservatism may mean that visualizations and interactions that guide the user toward partitioning data into smaller subsets or viewing multiple related visualizations at once, like trellis plots and visualization recommenders, are better for ensuring the analysts' inferences are appropriately sensitive to sample size than visualizations that encode many variables in a single view, assuming attention is not in scarce supply. 
These and other implications of Bayesian cognition and Bayesian model checking may lead to ideas for how to improve systems in ways that would be hard to predict without the theory. 

The primary challenge to integrating Bayesian cognition into the design and evaluation of interactive analysis tools is eliciting prior and posterior beliefs, as the method used to elicit priors influences the results (see \cite{o2006}, for a review) but it can be hard to evaluate whether one has gotten the right prior for a person. Validation approaches that present draws from the prior predictive distribution should help some here. The prior elicitation process may also shift any natural inference process, for example by causing a person to dwell more on their beliefs than they would. This may be useful or harmful depending on how much a user fixates on or overweights prior knowledge relative to that gleaned from the data. However, prior elicitation need not rely entirely on the user's ability to articulate expectations.  Many analysts do not work entirely on "one-off" data analyses but instead frequently analyze new data samples that use the same or similar data schema as previously analyzed data. Hence it may be possible for GUI tools to infer priors with more lightweight steering by the user by mining logs of prior database or data connection interactions on the part of a user.

\subsubsection*{Graphical inference as null hypothesis testing}
Some statisticians have proposed an analogy between graphical statistical inference and null hypothesis significance testing \citep{gelman2003,wickham2010}; \citet{buja2009} argue that discovering some insight using a visualization is akin to rejecting at least one assumption made under a null hypothesis. This understanding has led to several types of graphical tools.

The Rorschach method involves producing an array of ``null'' plots, visualizing data drawn from a null distribution that represents samples from a data generating process where no pattern exists~\citep{buja2009,wickham2010}. The idea is that looking at such plots can calibrate the eyes for sampling variation as one examines data.

\begin{figure*}
\centering
\includegraphics[width=\textwidth]{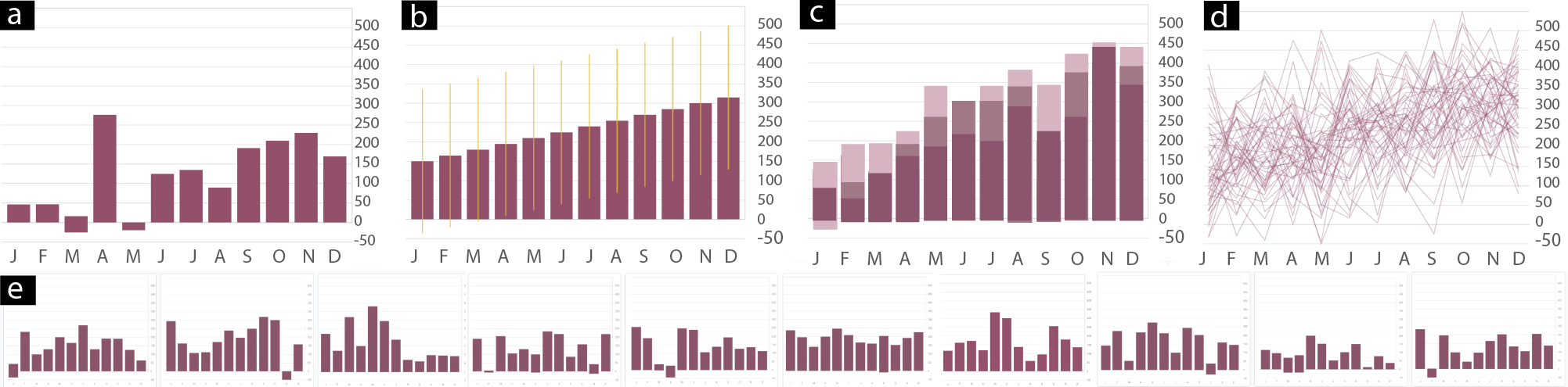}
 \caption{\it (a) Bar chart of (simulated) monthly estimates of jobs added or lost in the U.S. over one year. (b) Representation of a model that assumes steady monthly growth of 15k jobs with a standard deviation of 75k jobs. (c) Representation of the same growth model as animated hypothetical outcomes. (d) Fifty draws from the same model depicted as a static ensemble using lines instead of bars. (e) A lineup hides observed data among 9 null plots representing draws from a null model in which there is no growth. If users can identify the observed data from the lineup, they are said to have approximated a hypothesis test with type 1 error rate of $1/N$, in this case 0.10. 
 Data adapted from \citet{kale2018hypothetical}, which found in a within-subjects experiment that using animated hypothetical outcomes with a rate of 400ms per frame to display possible data generating processes led to more accurate judgments about the higher probability model than static ensemble or error bar representations.}
  \label{fig:hops}
\end{figure*}

The lineups approach also relies on null plots generated in the same way, but produces an array of $N$ plots, where one of the plots is of an observed dataset \textit{y} and the other $N-1$ plots are null plots~\citep{buja2009,wickham2010}. If an analyst can identify which of the $N$ plots shows the observed data, they are said to have performed a visual test equivalent to a hypothesis test with type 1 error rate of $1/N$. The lineup is demonstrated in Figure \ref{fig:hops}e, which hides a set of monthly jobs estimates among nine null plots drawn from a model with no growth. Both approaches require the analyst to specify the null generating mechanism. 



The Bayesian perspective on graphical comparisons as model checks subsumes treating visual comparisons as null hypothesis significance tests as a special case. The two frameworks align in many ways: both focus on the importance of judging deviation for some model assumptions, with the lineup and Rorschach representing general techniques for implementing graphical model checks.

Of course, some null mechanisms might be naive or obviously false~\citep{kale2019adaptation}, and having to specify the null mechanism adds a degree of freedom, so expecting inferences from lineups to be equivalent to doing an exact statistical test is problematic. We think it is unfortunate that lineups have been so strongly associated with statistical hypothesis testing, which may imply to non-statisticians like computer scientists that the technique is less about understanding deviation than it is about checking whether some difference is equal to zero, which is a priori implausible in many real world scenarios.
In their study of the multiple comparisons problem in exploratory analysis, \citet{zgraggen2018investigating} identified each analyst's explicit hypotheses (those stated by the participant) and implicit hypotheses (those not reported, but identified later in interviews or using eye tracking) to estimate how of many conclusions they drew were false positives. Since eye-tracking remains unrealistic to embed in real interactive analysis tools, other research proposes heuristics based on session logs to detect visual comparisons made while someone interacts with a visualization system \citep{zhao2017controlling}. For example, not every visualization with a filter is a hypothesis test, but every visualization with a filter condition is a test of the null hypothesis that the filter condition makes no difference compared to the distribution of the whole dataset. Such approaches imply a one-to-one mapping between graphical comparisons and statistical tests that is not well supported by the graphical inference literature. A graphical comparison supports the evaluation of many different ``tests'' simultaneously, some of which might not be well understood even by the analyst until they are violated. 
For example, even a simple two dimensional scatterplot can lead to a number of visual judgments of properties associated with scagnostics \citep{wilkinson2005graph}, like clumpiness, monotonocity, or skewness.

Recent research on lineups implies in various ways that the analogy between examining a lineup and doing an unbiased statistical test is not so simple, including the visual acuity of users and design of the visualization \citep{vanderplas2015spatial,vanderplas2017clusters}. Lineups have been applied, for example, to diagnose problems of fit with hierarchical models \citep{loy2017model} and to identifying the ``graphical power'' of different visualizations for supporting pattern finding \citep{hofmann2012graphical}, both use cases that are well aligned with the idea of graphical judgments as model checks more broadly. Recent work on dual lineups~\citep{vanderplas2017clusters}, which introduce two competing signals into a single lineup to identify which is more salient to a viewer, comes closer to a Bayesian comparison of the relative strength of two competing models with respect to a common null. In addition to demonstrating how the way visual encodings are used affects the viewer's chances of identifying one signal over another, this work also demonstrates how it can be difficult to define a null generating mechanism that matches the characteristics of non-target features. 
Studying how people look at lineups to better understand graphical statistical inference has become its own line of research \citep{beecham2016map,chowdhury2014utilizing,majumder2013,vanderplas2015spatial,zhao2013mind}, providing some evidence of our view above that a good attempt at formalizing graphical inference can lead to better understanding of human visual inference and where it deviates from expectations. 

Lineups can also be viewed as complementary to the Bayesian cognition approach summarized above, in that lineups fuse concepts from perceptual psychology like target identification and visual search with concepts from statistical modeling, while Bayesian cognition fuses concepts from cognitive psychology and behavioral economics like belief updating and revision with statistical modeling. 





\section{Implications for designing interactive analysis software}


\citet{tukey1986sunset} described how the development of expert systems helps address a challenge statisticians face in trying to teach statistical data analysis to the many people who need to use it. One reason is that ``[o]ne just cannot build an expert system without thinking through a strategy,'' hence designing a useful system prompts reflection on what a good strategy is. Another benefit is that a good expert system can be a way of teaching, such that ``[u]sing each well-planned system will then give continuing education-especially when the user repeatedly asks the system, `Why did it choose to do that?' After a while, some users will be ready for---nay will demand---more education, which we should by then be ready to furnish.'' 


There is an analogy between our vision for exploratory analysis software that more tightly integrates support for model-driven and probabilistic inferences and Tukey's observation that expert systems require reflection on strategies and pave the way for greater education on the part of their users. We suspect that the prioritization of pattern finding in current GUI tools for exploratory analysis is only partially intentional. Researchers may have gravitated toward optimizing for perception over cognition because it is easier to observe and thought to be less dependent on the data analysis context, leaving theories of human graphical inference underexplored. Researchers and developers of modern GUI systems consequently put analysts in an environment where they are encouraged to rely on their eyes' ability to perceive patterns and their intuitions about effective graphics, combined with the implicit guidance of system defaults. If an analyst wants to transition to more formal checks against candidate data generating processes, they must develop a process for doing so within the constraints of the tool, potentially sacrificing rigor when built-in functionality doesn't cover their modeling needs, or move to a different tool. 
This makes it harder for users to recognize where the patterns they perceive are tenuous, or even when they are making predictions versus exploring a space of theories. 

While we doubt that all users of visual analysis software assume implicit distributions when they judge ``signal'' in graphics, or even that very experienced users always consider distributions during visual analysis, evidence of the use of superficial visual heuristics for estimating effect size or pattern ``significance'' is not hard to find in our own and others' research  \citep{conti2005attacking,hofman2020visualizing,kale2020visual,nguyen2020exploring}.
Researchers and developers of interactive visual analysis tools should consider how software might encourage more robust inferences. This question presents an opportunity for thinking differently about how GUI EDA tools can support analysts' natural processes, but also a significant design challenge, since more integration with modeling may introduce the possibility of cognitive overload or misuses like overfitting.

\subsection{Design requirements and future directions}
At a high level, if model-driven inference underlies exploratory analysis, then systems should be capable of representing data generating processes. There are several functional implications of this. 

First, software should support and encourage the use of robust representations of uncertainty whenever inference may be a goal.
Above we discuss plotting observations rather than aggregations by default as one simple way software design can prioritize variation and uncertainty. However, implementing non-parametric bootstrapping as the basis for plotted data would take this a step further, potentially helping calibrate analysts to consider uncertainty by default while avoiding the need to train analysts on how to think about confidence intervals. How to provide the analyst control while encouraging them to contend with variation and uncertainty whenever patterns are being taken as ``findings'' is a question to be tackled in interaction design, and may require some diversity of strategies to be built in to GUI tools. 

\begin{wrapfigure}{r}{0.5\columnwidth}
  \centering
  \fbox{\includegraphics[width=0.5\textwidth]{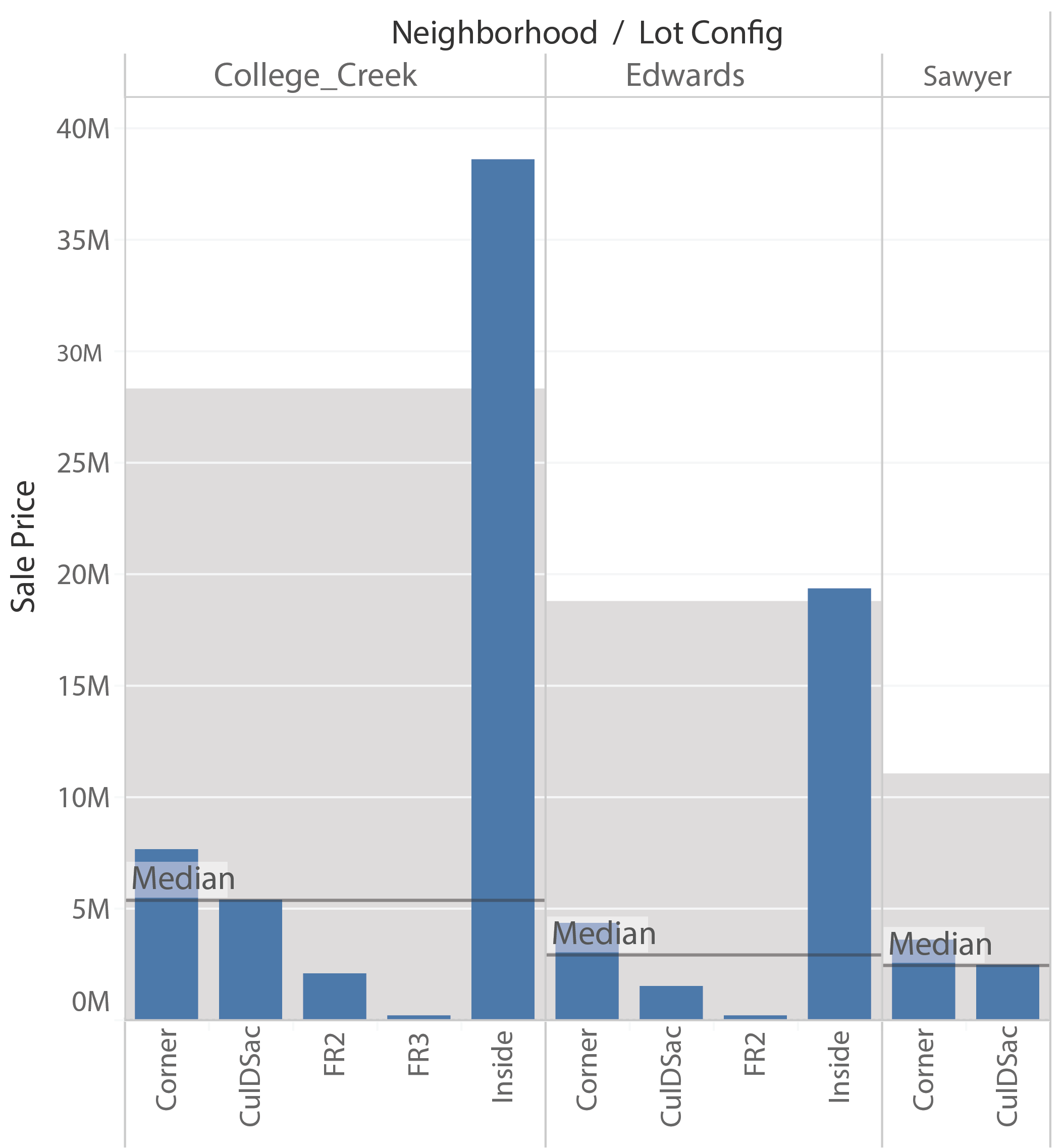}}
  \caption{Tableau's analytics pane supports the addition of arbitrary and data-based reference lines, along with standard uncertainty intervals based on the plotted data.}
  \label{fig:tableau_analytics}
\end{wrapfigure}

Second, analysts should have the ability to specify and see predictions from models of the data generating process that they wish to consider against the data.  
This idea is not entirely novel.  
Many widely used graphical user interface tools for data analysis, like Tableau, MS Excel, or Data Desk to name a few do provide modeling tools in the form of built-in statistical tests and regression features. In fact, Tableau Desktop, which we used to create most figures in this article, currently supports visualizing reference lines, bands, and empirical distributions~(\citeyear{tableaureferencelines}; Figure~\ref{fig:tableau_analytics}) as well as forecasting for time series data~(\citeyear{tableauforecasting}). However, these tools are intended for primarily confirmatory use or prediction based on the observed data, with little recourse to, for instance, customize based on a prior prediction or to easily compare different possible models of a data generating process.

Also related to our vision of supporting rough model checking are existing tools like the lineup and Rorschach \citep{wickham2010}, both of which have been proposed for comparing observed data to predictions of a null model. As currently implemented in R these tools require the analyst to specify the null model programmatically. What might it look like if these tools were implemented in GUI systems for exploratory analysis? 

Partly the difference is in emphasis. 
Systems could provide users with access to predictions of null models through built-in recommendations based on chart and data types. When visualizations include various distinct subsets of data, such as in trellis plots, the analyst could interactively select the data of interest. 
On the other side of the spectrum, analysts could see posterior predictive distributions from a model fit to data, using either a weakly informative or elicited prior, whenever visualizations of observed data alone seem ambiguous. When there is a clear source of prior information, for example as in business applications where similar analyses are conducted periodically on the latest sales or marketing data, seeing model predictions along with the observe data could help the analyst better perceive what if anything has been learned from the new information.

A key activity toward realizing this aim that has not been well explored by prior research is what a ``grammar'' for model recommendations should look like. At the minimum, such a grammar should include common distributional families like Gaussian, beta/binomial, Poisson, etc.; common transformations like taking a log; and support for simple additive and multiplicative models. Ideally there is a connection between the visualization structure and the model. The development of these tools should involve collaboration between computer scientists with expertise in designing interactions and accompanying abstractions and statisticians who bring expertise in robust statistical methods, such as models that minimize critical assumptions that can be fit using analytical solutions.

One direction for future research is to work toward developing a computational engine for automated model specification via user interactions with variables like dragging to shelves. Just like Tableau's underlying table algebra~(\cite{stolte2002polaris}) creates algebraic expressions from user-selected data fields, where fields are operands operated on by combination functions like cross, nest, and concatenate to drive visual specifications, a model-generating engine could automatically compute plausible regression models based on data characteristics. An analyst could visualize predictions of these models on demand, and customize the model specifications as needed by providing prior beliefs or varying assumptions.  
An engine might include precomputation of null alternatives as well. For example, showing draws from a model with non- or weakly informative prior fit to observed county data in a choropleth map depicting rates helps the analyst account for the impact of sample size in their inferences~(\cite{correll2016surprise}). 
Draws from a model that assumes all regions have the same rate, which might be set to the national average or some other baseline value that analyst decides, may also be useful. By using draws to represent the expected amount of sampling error at the sample size in each region, such models go beyond the canonical null model naive to population density. 
For greater flexibility, cases where users wish to target the modeling toward only some data in a view could directly select a subset by interacting with the chart, then right click to further parameterize the automatically computed model.  Feedback in terms of predictions from the model should be immediate and reversible, and the analyst should be able to control when and how they are shown (e.g., through animated draws~\citep{hullman2015hypothetical}, static ensembles, continuous representations, and either superimposed in the view or juxtaposed in separate views).





The Bayesian view of graphical examination as posterior predictive check motivates making it easier in tools for users to articulate prior distributions over parameters. Again in the interest of avoiding interrupting workflows to dive into code, the user could ideally `sketch' a prior graphically then draws from it and see these along with observed data. Thought should be put to how to do this, of course, given that different elicitation interfaces can lead to different strategies for formulating priors and modeling `noise' from the interface~(\cite{kim2019,sarma2020prior}).

Several existing research prototypes overlap with our proposal, in that they pursue a tighter integration with statistical modeling and aim to make statistical modeling more accessible to non-expert users of visual analysis tools. These include Statsplorer~\citep{wacharamanotham2015statsplorer}, which focuses on enabling users to select data in visualizations in order to run confirmatory tests directly, hence differing slightly from our proposal by shifting the focus toward confirmation over identifying deviation from model predictions, and Northstar~\citep{kraska2018northstar}, a pen-and-touch driven system that focuses on visualization combined with fast interactive machine learning.

Finally, on the flip side of the argument we make here, where confirmatory tools are currently integrated in GUI visualization tools, they should default to graphical model checks. Animated hypothetical outcomes \citep{hullman2015hypothetical} can be a useful tool here too, though the prevalence of visual impairments~\citep{chan2018estimates} motivates retaining access to table-based displays for viewers who need them. 

\subsection{Design challenges}
As a theory, by implying that software should allow and even encourage users to make reference distributions explicit, the model check formulation opens up room for considerably more complexity in GUI interfaces for exploratory analysis. This includes the formidable challenge of developing a grammar of flexible yet robust model specifications and ensuring that models can be fit when users customize them. Naturally there will be kinks in this process, making it important to invest in exploring various ways to give feedback to the analyst during model specification and exploration. 

One clear risk is that the additional cognitive load of interacting with reference distributions overwhelms some users, for example, distracting them from paying as much as attention to the data as they might have. 
A key question that often remains unstated in research in interactive visual analysis is: How much of the statistical inference process that an analyst engages in should be left implicit in order to preserve cognitive load? We acknowledge that it is difficult to answer this question without first making concerted attempts in research to realize the forms of integration we describe above. However, as we argue above, there are many reasons to explore more explicit integration of statistical modeling in exploratory visual data analysis tools.

An user-centered design approach toward prototyping and evaluating the use of different interaction designs for modelling tools will be necessary along with the development of a grammar of model components. These efforts should work toward guiding design principles, some of which may be similar to those used for mixed initiative interface design \citep{horvitz1999principles}, like allowing direct invocation and termination and providing dialogue to resolve key uncertainties rather than making guesses that a user may not realize were made. A user-centered design process should be followed by user testing with business, public sector, and academic populations, where user samples are representative of the range of skill sets in statistical graphics and modeling in those sectors.

Another risk is that adding support for reference distributions introduces new ``failure modes'' based on misunderstandings of how the features are meant to be used, or failures of the system designers to anticipate what details of data structure will be critical to infer or elicit. In particular, if we expect to encounter tradeoffs in how easily implementable a model is and how appropriate it is for the sorts of real world scenarios analysts bring to GUI tools, then we risk leading analysts to overrely on inappropriate models. If analysts began to respond to the tools more than the data, the link between the models they use and their intuitive theories is weakened, which might lead to analyses that are less responsive to the data. 
Overfitting is another concern. More built in functions for generating model-driven expectations from existing data may, if not designed to promote debugging and skepticism, exacerbate issues. This possibility motivates design features for separating some data for testing when analysts want to check any hypotheses they arrive at. More broadly, a challenge is how EDA tools can encourage the use of modeling for building understanding of data generating processes, accounting for uncertainty, and making predictions that can be tested on future data over manufacturing provable insights. All of these challenges are formidable, but without trying to push the horizon of what the average GUI tool supports in terms of support for model checking, it is difficult to say how limiting they will be. Again iterative user-centered in the development of these tools should help identify how failures can occur and why early in the design process.

At a higher level, we expect the potential complexity of supporting reference model comparisons in GUI analysis tools may seem unwelcome to many who have long accepted the model-free or ``leave it all implicit'' philosophy in interactive analysis system design. Certainly these activities should be approached cautiously; visual analytics software has been criticized for unwarranted complexity in the form of overloading an interface with available functions \citep{hegarty2010visweek}.  
However, rather than implying that complex modeling must accompany all graphical examinations, we think the model check theory is better seen as an opportunity for those developing systems to reflect more deeply on plotting or other strategies analysts may currently use to help them make model driven judgments. This allows developers to target modeling tools to cases where the status quo of implicit model checking with graphics is most likely to be prone to vague or erroneous understandings of the data on the part of the analyst, since here there is more to gain. In particular, users of interactive visual analysis tools who lack substantial statistical or analytical background might be most prone to failing to realize where their attempts to find patterns are tenuous or how they could benefit from thinking about the underlying process producing data. 

Finally, we acknowledge that we might be wrong, and attempts to more tightly integrate modeling into default modes of interactive visual analysis tools might overwhelm most analysts or lead to more brittle interpretations than the current status quo supports. However, we still see theory as unavoidable for continued progress in research and development. In particular, more careful consideration and formal reasoning about the  problem space that visual analysis is intended to address---from signal so large that it ``hits you between the eyes'' to patterns so overwhelmed by noise and confounding that further statistical modeling would be useless---may help researchers better address some of the contradictions posed by critical empirical work.

\section{Comparing human to automated statistics}
Beyond the design implications of a Bayesian model check formulation, there are various testable expectations about human graphical inference that could lead to a better understanding of how people do exploratory visual analysis. 

For example, how ubiquitous is conservatism in belief updating, as suggested by recent work in behavioral economics (e.g., \cite{benjamin2016}) and visualization (e.g., \cite{kim2019})?  
What predicts the use of non-probabilistic pattern detection activities for theory or model exploration versus implicit graphical model checking in an analysis session? And how do changes to graphical representations (such as showing animated bootstrap replicates) affect visual analysis? Some of this work will naturally help researchers see what can't be well identified about intuitive graphical inference. For example, to faithfully infer implicit reference distributions from experiments on human graphical inference could require more modeling structure of correlations than would be reasonable to assume in viewers' conscious reasoning processes. But the act of trying to specify users' processes statistically leaves us with a better understanding of what we can presume about exploratory analysis versus what remains subject to conjecture.

We think the research trend toward studying how data analysts use existing GUI systems (e.g., \cite{battle2019characterizing}) is well motivated and much more could be done under the umbrella of deepening a theoretical foundation.
For example, what do different classes of inference look like in expert use of visual analysis tools? How do analysts use interactive graphics when considering different causal hypotheses, and can we learn anything about human graphical causal inference by comparing to models of causal support from mathematical psychology \citep{tenenbaum2001structure}?


Finally, beyond using theories to produce testable statements about how humans do analysis and to stimulate new design ideas, considering how we might remove humans from the analysis process entirely may also paradoxically help us find ways to improve interactive analysis. In other words, how could an Artificial Intelligence do statistics? We use this question as a thought exercise for further reflecting on the types of knowledge and strategies that come into play during interactive analysis.

In the old-fashioned view of Bayesian data analysis as inference-within-a-supermodel, it's simple enough to imagine an AI replacing a person: it simply runs some equivalent to a probabilistic program to learn from the data and make predictions as necessary. But in a modern view of statistical practice---iterating the steps of model-building, inference-within-a-model, and model-checking---it's not quite as clear how the AI works. By taking what currently seems vague and framing it computationally, we might discover useful regularities or patterns in human statistical workflows.


To fix ideas, we shall discuss Bayesian data analysis, which can be idealized by dividing it into the following three steps \citep{gelman2013bda}:

\begin{enumerate}
\item Setting up a full probability model—a joint probability distribution for all observable and unobservable quantities in a problem. The model should be consistent with knowledge about the underlying scientific problem and the data collection process.

\item Conditioning on observed data: calculating and interpreting the appropriate posterior distribution—the conditional probability distribution of the unobserved quantities of ultimate interest, given the observed data.

\item Evaluating the fit of the model and the implications of the resulting posterior distribution: how well does the model fit the data, are the substantive conclusions reasonable, and how sensitive are the results to the modeling assumptions in step 1? In response, one can alter or expand the model and repeat the three steps.
\end{enumerate}
Currently, human involvement is needed in all three steps listed above, but in different amounts:

\begin{enumerate}
\item Setting up the model involves a mix of look-up and creativity. We typically pick from some conventional menu of models (linear regressions, generalized linear models, survival analysis, Gaussian processes, splines, trees, and so forth). Machine learning toolboxes and probabilistic programming languages such as Stan enable putting these pieces together in unlimited ways, with similar expressiveness to how we formulate paragraphs by putting together words and sentences. Right now, a lot of human effort is needed to set up models in real problems, but we could imagine an automatic process that constructs models from parts. 

\item Inference given the model is the most nearly automated part of data analysis. Model-fitting programs still need a bit of hand-holding for anything but the simplest problems, but it seems reasonable to assume that the scope of the ``self-driving inference program'' will gradually increase. For example, for thirty years we have been able to automatically monitor the convergence of iterative simulations \citep{gelman1992}. With the no-U-turn sampler, a recursive algorithm builds a set of likely candidate points spanning a wide swath of the target distribution and stops when it starts to retrace its steps, thus avoiding the need to tune the number of steps in Hamiltonian Monte Carlo \citep{hoffman2014}.

\item The third step---identifying model misfit and, in response, figuring out how to improve the model---is likely the toughest part to automate. We often learn of model problems through open-ended exploratory data analysis, where we look at data to find unexpected patterns and compare inferences to our statistical experience and subject-matter knowledge. Indeed, a primary piece of advice we espouse to statisticians is to integrate that knowledge into statistical analysis, both in the form of formal prior distributions and in a willingness to carefully interrogate the implications of fitted models.
\end{enumerate}


By considering how to fully automate all three steps, we can identify some ways to improve interactive software. The space of model parts we deem necessary to support step 1, for example, should directly guide the types of built-in options that interactive analysis tools offer an analyst to specify their implicit models. 
When it comes to step 2, inference within the model, we might try to build in automatic checks (for example, based on adaptive fake-data simulations) to flag problems with fitting a specified model when they appear. This could help us think about how users might note immediate problems with an implicit model as they examine graphics.

How would an AI do step 3? The AutoML approach to model evaluation typically involves choosing a preferred loss function to minimize, e.g., generalization error on held-out data, estimated using standard procedures like cross validation. But human model checking often combines model fitness measures with more qualitative assessments of how well model predictions align with domain knowledge. One approach closer to human model checking is to simulate the human in the loop by explicitly building a model-checking module that takes the fitted model, uses it to make all sorts of predictions, and then checks this against some database of subject-matter information such as a knowledge graph. This is one avenue for attempting to mimic the Aha process behind concepts like insight that drives scientific revolutions. Trying to construct this would undoubtedly require deeper inquiry into how humans check model fit, and might lead to ideas for building interactive systems, like making it easy for analysts to scan through many predictions from their models or transform them into different measures to ask ``does this look right.''

All that is left, then, is the idea of a separate module that identifies problems with model fit based on comparisons of model inferences to data and prior information. It's less clear how techniques from AI and ML research should be combined to do that; this may be the hardest part of the pipeline to remove humans from the loop. However, by attempting to combine existing technologies we are likely to learn more about how to think about humans doing model checks, which might also feed new interface optimizations.



\section{Conclusion}


Data visualization and exploratory data analysis can be seen as a form of model checking, with the goal of revealing the unexpected beyond what is already in a model of the world. 
We propose a research program that pursues a tighter integration between models, graphics, and data querying, motivated by a view of interactive analysis as a process of users comparing intuitive pseudo-statistical models to data via model checks. A Bayesian model check formulation of exploratory visual analysis makes clearer what types of interactive features would help in phases of analysis that resemble rough CDA, like built-in reference distributions, robust uncertainty representations, and features to encourage analysts to recognize the links between assumed models and priors and graphical structures. 

Our proposal of Bayesian model checking as a way to unify thinking about different phases of analysis calls for more thoughtful integration of graphical inference techniques that researchers are already proposing into visual analysis systems. This includes innovations in uncertainty visualization like animated and static frequency framed depictions of distributions and variations on graphical inference techniques like lineups and graphical elicitation of priors or predictions from users, to name a few. While its natural to expect that novel techniques will not necessarily immediately make their way into mainstream tools, we suspect that a relative lack of theory around what visual analysis should be stands in the way of graphical inference tools becoming a standard part of the visual analysis toolkit.

Beyond stimulating new ideas for designing interfaces and new ways of evaluating them, conceiving of interactive analysis as checks against pseudo-statistical mental models pushes us toward identifying testable implications of different formalizations of this process. 
Without a theoretical foundation fields like interactive visualization and visual analytics can become trapped in a problem-solving orientation to system development. In such an orientation, researchers may be more likely to continually chase the next application area to design for than to consider ways the design of an interactive system might help users recognize their own goals and limitations. 
A good theoretical framework of modeling feeds a process in which we learn from the ways that peoples' behavior deviate from model predictions and continually revise our aims as researchers and developers, including our understanding of the problem space we aim to address with software. 
This process is not completely absent in the status quo approach: current efficiency-oriented evaluations of interactive systems can also help researchers realize when their intuitions are wrong. The point it is that its likely to be less direct and error prone than if a more formal, normative model were available, similar to how it is inefficient to learn from only the yes/no answers of null hypothesis significance tests.


Our argument about the potential consequences of prioritizing pattern exposure in creating tools for interactive EDA should not be construed as saying that exposing raw data is generally bad in analysis contexts. 
In contexts like communicating statistical results, showing the data or properties of the raw data can be very useful for providing information about effect size, especially in light of many readers' tendencies to overestimate effects \citep{hofman2020visualizing}. 
And, as system developers and researchers, we face many daunting challenges that existing pattern-focused tools address well. For example, to make huge datasets interactive at all requires a number of database and visualization-based optimizations.  Similarly many of the interactive analysis innovations we've surveyed, such as recommendations based on graphical (e.g., \cite{wongsuphasawat2015voyager,wongsuphasawat2017voyager}) or statistical features (e.g., \cite{lee2018case,vartak2015s}) have an important role to play in reducing the many manual efforts required to do interactive analysis. 
However, we think the field of interactive data analysis could better achieve its goals of transforming how people interact with data if such innovations were guided by theories of inference.
This is not to suggest that this task will be easy, as there is much still to learn about how to gently introduce modeling capabilities without interrupting an analyst's flow, and about what users of different profiles do given more advanced modeling tools and asked to specify their expectations. 


We have focused our discussion primarily on analysis applications involving abstract data, where standard statistical graphics are the norm. In some other applications of interactive analysis and scientific visualization, users may have a harder time expressing their implicit models. For example, doctors might want to search for clinical features in large databases of medical imagery to help them in making diagnoses. When experts' implicit models are based in recognizing of visual-spatial signatures, it may be harder to elicit them, or at least require very different interfaces than we propose here. However, the fact that interactive interfaces are moving toward eliciting more input from domain experts like doctors' to facilitate their work even when their implicit models are hard to formally represent suggests some parallels despite the different assumptions that can be made about the data \citep{cai2019human}.


Finally, a good model of intuitive graphical inference is likely to have implications for communicative use of interactive and static visualization as well. We have used the Bayesian model checking formulation to theorize about the role played by uncertainty communication in communicative visualization, for example~\citep{hullman2019}. Visualizations are sometimes described as storytelling devices. The connection here is that stories can themselves be viewed as model checks or as explorations of anomalies, with the ``twist'' in a good story corresponding to a confounding of expectations \citep{gelman2014stories}. Putting these together suggests that designers and readers should consider visualizations with respect to the expectations or default narratives they overturn. A good model of inference can help us see the similarities between more than one pair of seemingly opposed activities.

\subsection*{Disclosure Statement}
The authors have no conflicts of interest to declare.

\subsection*{Acknowledgments}
 We thank the National Science Foundation for grant 1749266, Microsoft, the U.S. Office of Naval Research for grant N000141912204, the Institute for Education Sciences for grant R305D190048, and Alex Kale, Jeffrey Heer, Hanspeter Pfister, Matthew Kay, Jennie Rogers, and the anonymous reviewers for helpful comments.
 

\appendix


\printbibliography

\end{document}